\documentclass[preprint,showpacs,preprintnumbers,amsmath,amssymb,nofootinbib]{revtex4}
\usepackage{graphicx,color,epsfig}

\begin{document}

\title{New method for extracting quasi-bound states from the continuum}

\author{J.B. Faes and M. P{\l}oszajczak}
\address{Grand Acc\'{e}l\'{e}rateur National d'Ions Lourds (GANIL),CEA/DSM -- 
CNRS/IN2P3, BP 55027, F-14076 Caen Cedex 05, France}


\begin{abstract}
A new parameter-free method is proposed for treatment of single-particle resonances in the real-energy continuum shell model. This method yields quasi-bound states embedded in the continuum which provide a natural generalization of  weakly bound single-particle states.
\end{abstract}


\maketitle
\section{Introduction}\label{intro}

Many-body states of nuclear shell  model (SM)
are linear combinations of Slater determinants built of bound single-particle (s.p.) states. Thus, the SM describes bound many-body systems which are isolated from the external environment of scattering states and decay channels. A natural generalization of the SM for weakly bound or unbound many-body systems, the so-called  Gamow shell model (GSM), has been formulated recently in Berggren ensemble consisting of bound s.p. states, s.p. resonant states (Gamow \cite{Gam28} or Siegert \cite{Sie39} states), and the complex-energy s.p. continuum states \cite{Mic02a,Bet02}. A complete set of s.p. states can be defined in the Berggren ensemble \cite{Ber68,Ber93}. The complete set of many-body states is then given by all Slater determinants spanned by s.p. states of the complete s.p. set in the Berggren ensemble \cite{Mic02a,Bet02}. So defined theoretical framework gives the full description of interplay between scattering states, resonances and bound states in the many-body wave function, without imposing a limit on the number of particles in the scattering continuum. On the other hand, the asymptotic decay channels in GSM are not individually resolved.  Hence, this approach cannot be applied for a description of nuclear reactions and remains the tool for nuclear structure studies.

The unification of structure and reaction theories is possible in the continuum shell model (CSM) formalism \cite{Mah69,Bar77,Phi77}, including the recently developed shell model embedded in the continuum (SMEC) \cite{Ben99,Ben00,Oko03,Rot05}. In this formalism, the s.p. basis includes bound states and the real-energy continuum states. Feshbach's projection technique \cite{Fes58}, used in 
CSM and SMEC, allows to describe on the same footing the nuclear reactions, including the rearrangement term, and the nuclear structure of well-bound, weakly-bound or unbound nuclear states. 

Bound and scattering s.p. states define two orthogonal subspaces  $q_0$ and $p_0$, respectively. The s.p. resonances do not belong to the Hilbert space and have to be regularized before including them in the SMEC framework. The regularization procedure consists of  including s.p. resonances in a discrete part of the spectrum after removing the scattering tails which are included in the embedding scattering continuum. These regularized resonances are usually called the quasi-bound states embedded in the continuum (QBSEC). The new subspaces: $q$ including bound  and QBSEC s.p. states, and $p$ including non-resonant scattering states and scattering tails of regularized resonances, are subsequently reorthogonalized.  

The many-body states in SMEC are given by all Slater determinants spanned by s.p. states in $q$ and $p$. Many-body states with all particles occupying bound and QBSEC s.p. states span the $\mathcal{Q} \equiv \mathcal{Q}_0$ subspace of the Hilbert space. The complement subspace $\mathcal{P}$ includes many-body states with one or more particles in the scattering states, hence:  $\mathcal{P} \equiv \sum_{i=1}^{A}\mathcal{Q}_i$, where $\mathcal{Q}_{i}$ 
projects on the space with $i$ particles in the continuum. 

The number of particles in the scattering continuum provides a natural hierarchy of approximations  in the CSM.  Technical complications associated with inclusion of multiparticle continua and complex asymptotic channels are such that none of the early CSM or SMEC studies considered more than one particle in the continuum. Only recently, two-particle continuum has been included in the SMEC for the description of the two-proton radioactivity \cite{Rot05}. The general framework unifying the reaction theory and the structure theory for problems with any number of particles in the scattering continuum have been worked out as well \cite{Fae07}. 

One of key elements in the practical implementation of SMEC scheme is the definition of orthogonal subspaces $q$, $p$ and, consequently, $\mathcal{Q}_0, \mathcal{Q}_1, \mathcal{Q}_2, \dots$  many-body subspaces. This definition is associated with the extraction of regularized resonances from the s.p. scattering continuum. In all previous SMEC applications (for an extensive list of applications see Refs. \cite{Ben99,Ben00,Oko03,Rot05}) as well as in CSM studies of Dresden group \cite{Bar77,Rot91}, the Wang and Shakin method was employed \cite{Wan70}. In this method, the construction of a QBSEC depends on few parameters: the (real) energy of a scattering wave function which is then regularized and removed from the continuum, and the parameters of the cutting function (Heaviside or Fermi functions) chopping off the tail of this wave function. The so defined QBSECs are auxiliary, artificial objects which do not correspond to any solution of the Schr\"{o}dinger equation.

Unfortunately, the CSM/SMEC results concerning, e.g., the partial decay widths, depend on the parameters of the cutting function. 
This ambiguity cannot be totally removed even if the condition on the s.p. width is applied in choosing the cutting radius \cite{Bar77}. In practical applications, the radius of the cutting function is selected close to the top of the  Coulomb barrier \cite{Oko03,Rot05} what yields a sensible prescription for narrow s.p. resonances. Problems with the choice of cutting function and the extraction of QBSECs appear for broad resonances. Moreover, the QBSECs obtained using Wang and Shakin method \cite{Wan70} do not have the correct asymptotic behavior.

In this paper, we present the new method of regularizing s.p. resonances which is unambiguous, parameter-free and yields states with the correct bound-state asymptotics. These bound states embedded in the non-resonant continuum are called the {\em anamneses} of resonances in the space of square-integrable functions 9${\cal L}^2$-functions0. 

In Sect. \ref{radicons}, we provide a short discussion of different radial solutions of the Schr\"odinger equation which are used in the construction of a complete s.p. basis in SMEC. 
Qn unambiguous determination of the resonance anamneses is presented in Sect. \ref{new_qbsec}, and 
Sect. \ref{non_res_cont} is devoted to the discussion of consequences of the extraction of resonance anamneses on the non-resonant, real-energy continuum states. Together, bound s.p. states, anamneses of s.p. resonances and a regularized real-energy s.p. scattering states provide the complete s.p. basis in the Hilbert space. This basis can be used to obtain the complete many-body basis in CSM/SMEC studies.

Applications of the new method are discussed in Sect. \ref{applications}. We shall present examples of the anamneses for resonances of different widths and angular momenta $\ell$, both for neutrons and protons. General features of the resonance anamneses, such as the energy dependence of both the root-mean-square (RMS) radius or the matching point between inner and outer solutions of the Schr\"{o}dinger solution, will be analyzed as well. Finally, main conclusions will be given in Sect. \ref{conclusion}.

\section{Radial Schr\"odinger equation: the general considerations}
\label{radicons}

Let us consider a spherical potential $V(r)$ describing qn interaction between a nucleon 
and a target nucleus. In the center of mass coordinates, the one-body radial wave function 
$u(r)$ of the relative motion is the solution of the Schr\"odinger equation: 
\begin{eqnarray}\label{local_schr}
\left[ \frac{d^{2}}{dr^{2}}+k^{2}-\frac{\ell(\ell+1)}{r^{2}}-\frac{2\mu}{\hbar^{2}}V(r)\right]u_{k,\ell}(r) = 0 ~ \ , 
\end{eqnarray}
where $\mu$ is the reduced mass  and $\ell$ is the relative angular momentum. In the following, we shall omit the angular momentum index $\ell$ to simplify notations.

We are  interested in three kinds of solutions of Eq. (\ref{local_schr}): 
\begin{itemize}
\item The scattering solutions, which form a continuum for real and positive momenta $k$.
These solutions are regular at $r=0$ and asymptotically take a form: 
\begin{eqnarray}\label{asym_scatt}
u_{k}(r) &\sim& kr\Big{(}C^{-}h^{-}(kr)+C^{+}h^{+}(kr)\Big{)} ~ \ , 
\end{eqnarray}
where $h^{\pm}$ are irregular Coulomb wave functions or Hankel functions for protons and neutrons, respectively. $C^{\pm}$ in (\ref{asym_scatt}) are some constants. 
\item The bound state solutions, which compose a discrete set for imaginary and positive values 
$k=k_{n}$. These solutions have an asymptotic of outgoing waves: 
\begin{eqnarray}\label{asym_bound}
u_{n}(r) \sim Ak_{n}rh^{+}(k_{n}r) ~ \ .
\end{eqnarray}
\item The resonance solutions, which constitute a discrete set for $k=k_{n}^{res}$ such that 
${\cal R}e(k_{n}^{res})>0$, ${\cal I}m(k_{n}^{res}<0)$ and ${\cal R}e(k_{n}^{res})>-{\cal I}m(k_{n}^{res})$. These 
solutions have also an outgoing wave asymptotic. 
\end{itemize}
The asymptotic form  of irregular Coulomb wave functions (Eqs. (\ref{asym_scatt}), (\ref{asym_bound})) is:
\begin{eqnarray}\label{coul_conv}
h^{\pm}(kr)\sim \frac{i^{\mp(\ell+1)}}{kr}e^{\pm i(kr-\eta\log(2kr))} ~ \ , 
\end{eqnarray}
where $\eta=\mu e^2Z_1Z_2/{\hbar}^2k$ is the Sommerfeld parameter. 

The numerical integration for these three kinds of solution is carried out first by integrating the regular solution in the inner region  $[0,R]$, where $R$ is some matching radius. On this interval, the solution of 
Eq. (\ref{local_schr}) is written as: 
\begin{eqnarray}\label{reg_sol}
u(r)=Cu^{reg}_{k}(r)~ \ , 
\end{eqnarray}
where the regular solution $u^{reg}_{k}$ verifies: 
\begin{eqnarray}\label{reg_at_zero}
\lim_{r\to 0}r^{-\ell-1}u^{reg}_{k}(r)=1 ~ \ , 
\end{eqnarray}
and $C$ is a constant. Then, we integrate the Jost functions $H_{k}^{\pm}$ which are the
solutions of  Eq. (\ref{local_schr}) with the asymptotic form when $r\to \infty$:
\begin{eqnarray}\label{asym_jost}
H_{k}^{\pm}(r)\sim krh^{\pm}(kr) ~ \ . 
\end{eqnarray}
In the outer region $[R,\infty[$, the scattering solutions are: 
\begin{eqnarray}\label{scatt_jost}
u_{k}(r)=C^{-}H_{k}^{-}(r)+C^{+}H_{k}^{+}(r)~ \ .
\end{eqnarray}
The continuity of the logarithmic derivative, which ensures the matching between the two solutions
(Eqs. (\ref{reg_sol}) and (\ref{scatt_jost})), gives two conditions: 
\begin{eqnarray}\label{set_equa}
Cu^{reg}_{k}(R) &=& C^{-}H_{k}^{-}(R) + C^{+}H_{k}^{+}(R) \nonumber \\ \\
C\frac{du^{reg}_{k}(r)}{dr}\Big{|}_{r=R} &=& C^{-}\frac{dH_{k}^{-}(r)}{dr}\Big{|}_{r=R} + 
C^{+}\frac{dH_{k}^{+}(r)}{dr}\Big{|}_{r=R}  \nonumber
\end{eqnarray}
for three unknown constants $C^{\pm}$ and $C$. The third equation is provided by the normalization condition for scattering states: 
\begin{eqnarray}\label{C+C-}
C^{+}C^{-}=\frac{1}{2\pi} ~ \ . 
\end{eqnarray}

On the interval $[R,\infty[$, bound and resonance wave functions are written as: 
\begin{eqnarray}\label{bound_jost}
u(r)=AH^{+}_{k}(r) ~ \ . 
\end{eqnarray}
The matching between the internal (\ref{reg_sol}) and external (\ref{bound_jost}) solutions  is ensured for those discrete values of $k_{n}$ which nullify Wronskian of the regular solution with the Jost solution  $H^{+}_{k}$: 
\begin{eqnarray}\label{wron_equa}
W(u^{reg}_{k},H_{k}^{+})\Big{|}_{k=k_{n}}(r)\equiv0 ~ \ . 
\end{eqnarray}
 For resonance states, the values of $k_n$ correspond to the poles of the scattering matrix ($S$-matrix),  whose matrix elements are given by the ratio of outgoing and incoming waves in the asymptotic expression (\ref{asym_scatt}).

Properties of the scattering continuum are very sensitive to the position of 
these poles in the complex $k$-plane. For a narrow resonance, the scattering states with $k$-values on the real axis just above the corresponding pole of the $S$-matrix are localized in the inner region of a potential and resemble bound states with, however, an oscillatory asymptotics instead of an exponential asymptotics expected for standard bound state wave functions. This generic feature of near-pole scattering wave functions helps  to define the QBSEC function by cutting off the oscillatory tail with either the Heaviside function \cite{Wan70} or the Fermi \cite{Oko03} function.

\section{Determination of the anamneses of s.p. resonances}
\label{new_qbsec}

The significance of a QBSEC function in the CSM/SMEC theoretical framework is due to the similarity between a near-pole scattering wave function in the inner region ($r<R$) and a bound state wave function. The behavior of QBSEC in the outer region ($r>R$) does not correspond to any solution of the Schr\"{o}dinger equation and, in this sense, can be considered unphysical. In the following, we shall show that one can define an alternative to QBSEC functions, the anamneses of s.p. resonances which belong to the space of ${\cal L}^2$-functions. These resonance anamneses are bound s.p. states in the scattering continuum. They provide a natural continuation of weakly bound s.p. states for positive energies and extract all resonance features from the scattering continuum. 

Given a pole of the $S$-matrix at $k^{res}$ in the complex $k$-plane, this similarity is particularly striking for scattering states with $k$ close to:
\begin{eqnarray}\label{value_kappa}
\kappa=\sqrt{{\cal R}e\Big{(}(k^{res})^{2}\Big{)}} ~ \ . 
\end{eqnarray}
It is then quite natural to select the scattering state with a real eigenenergy 
$e^{res}=\hbar^{2}\kappa^{2}/2\mu$ for a construction of the resonance anamnesis.

In the inner region $[0,R]$, where $R$  is yet arbitrary matching radius for inner and outer solutions, the resonance anamnesis  should be proportional to the regular solution of Eq. (\ref{local_schr})  
with $k=\kappa$:
\begin{eqnarray}\label{qbsec_reg}
v_{\kappa}(r)=Cu^{reg}_{\kappa}(r) ~ \ , 
\end{eqnarray}
where $C$ is some constant. 

In the external region ($[R,\infty[$), we require
that the anamnesis of the resonance has the bound state asymptotic corresponding to $k=i\kappa$. 
Using the Jost solution $H^{+}_{i\kappa}$ of Eq. (\ref{local_schr}) with $k=i\kappa$, 
we write the wave function of resonance anamnesis in the interval $[R,\infty[$ as:
\begin{eqnarray}\label{qbsec_jost}
v_{\kappa}(r)=AH^{+}_{i\kappa}(r) ~ \ , 
\end{eqnarray}
where $A$ is some constant. Since the wave number is fixed, both in the inner $[0,R]$ and outer 
$[R,\infty[$ regions, therefore the continuity of the resonance-anamnesis wave function and its first derivative at $r=R$ is equivalent to the condition:
\begin{eqnarray}\label{qbsec_cont}
W(u^{reg}_{\kappa},H^{+}_{i\kappa})(r)\Big{|}_{r=R}=0~ \ . 
\end{eqnarray}
Since the wave number is fixed, both in the inner $[0,R]$ and outer $[R,\infty[$ regions, 
Eq. (\ref{qbsec_cont}) becomes a condition fixing the value of the matching radius
$r_{m}\equiv R$. Hence, the wave function of the resonance anamnesis on $[0,\infty[$ can be as:
\begin{eqnarray}\label{qbsec_wf}
v_{\kappa}(r)=C\left[u^{reg}_{\kappa}(r)
\Big{(}1-\Theta(r-r_{m})\Big{)}+\frac{u^{reg}_{\kappa}(r_{m})}{H^{+}_{i\kappa}(r_{m})}H^{+}_{i\kappa}(r)\Theta(r-r_{m})\right]_{r_m=R} ~ \ . \nonumber \\
\end{eqnarray}
$\Theta$ in this equation denotes the Heaviside function and the constant $C$ is fixed by the normalization condition for $v_{\kappa}(r)$.

The anamnesis of a s.p. resonance should be orthogonal to other bound and scattering states of a given $\tau_z,\ell,j$. If a bound state exists with the same angular momentum $\ell,j$, then the wave function of the resonance anamnesis $v_{\kappa}(r)$ (cf Eq. \ref{qbsec_wf}) has to be orthonormalized with respect to this state. Construction of an appropriate non-resonant scattering continuum will be discussed in Sect. \ref{non_res_cont}.

The above procedure yields for the resonance anamnesis a ${\cal C}(1)$-function with an appropriate bound state asymptotic. It is important to notice that the radius $r_{m}(\equiv R)$ in (\ref{qbsec_wf}) is 
determined by solving  Eq. (\ref{qbsec_cont}), and does not correspond to any arbitrary cutting radius as in the procedure of Wang and Shakin \cite{Wan70}.  We will see in Sect. \ref{applications} that despite the fact that $r_m$ can in principle take any value in the interval $[0,\infty[$,  the above method provides the resonance anamneses which maintain localized aspects of the corresponding resonance wave functions. 

\section{Construction of the non-resonant continuum}
\label{non_res_cont}
The radial wave functions corresponding to resonance anamneses are not eigenfunctions of the original Hamiltonian ${\hat h}$ which is used to define s.p. bound and scattering states. It is however possible to redefine this Hamiltonian in such a way that bound states of ${\hat h}$, resonance anamneses,  and the states of a non-resonant scattering continuum are generated by one and the same Hamilton operator. 

Let us denote by $\{|u_{n}\rangle\}$ the state vectors corresponding to bound states, and by $\{|v_{m}\rangle\}$ the state vectors corresponding to the resonance anamneses defined in the previous section. The energy of each resonance anamnesis corresponds to the resonance energy:
$e^{res}_{m}=\hbar^{2}\kappa_{m}^{2}/2\mu$, where $\kappa_{m}$ is given by (\ref{value_kappa}), and 
an index $m$ enumerates different poles. Bound states and resonance anamneses form together a discrete subset $\{|\tilde{u}_{n}\rangle\}$ of the complete set of basis states in Hilbert space  \cite{New82,Mic04}. 

Let us now define a new Hamilton operator ${\hat {\tilde{h}}}$: 
\begin{equation}\label{new_ham}
\hat{\tilde{h}}=\sum_{n}|\tilde{u}_{n}\rangle\tilde{e}_{n}\langle\tilde{u}_{n}|+\hat{p}{\hat h}\hat{p}~ \ , 
\end{equation}
where $\hat{p}$ is a projection operator on the non-resonant s.p. continuum:
\begin{equation}\label{projector}
\hat{p}=1-\sum_{n}|\tilde{u}_{n}\rangle\langle\tilde{u}_{n}| ~ \ , 
\end{equation}
and
\begin{equation}\label{new_energies}
\tilde{e}_{n} = 
\left\{
\begin{array}{ll}
e_{n} & \rm{for~a~bound~state} \\[+0.3cm] 
e^{res}_{n} & \rm{for~an~anamnesis~of~the~resonance~state} 
\end{array}
\right. 
\end{equation}
All vectors in the set $\{|\tilde{u}_{n}\rangle\}$ are the eigenstates of $\hat{\tilde{h}}$: 
\begin{equation}\label{new_ham_equa}
(\tilde{e}_{n}-\hat{\tilde{h}})|\tilde{u}_{n}\rangle=0 ~ \ .
\end{equation}
The scattering continuum is renormalized by the second term on the r.h.s. of Eq. (\ref{new_ham}). 
Below, we shall discuss the construction of scattering states for the modified Hamiltonian 
$\hat{\tilde{h}}$ (cf Eq. (\ref{new_ham})). Such states, that we denote $|\tilde{u}\rangle$, have to be solutions of the equation:
\begin{equation}\label{a1}
(\tilde{e}-\hat{p}\hat{h}\hat{p})|{\tilde u}\rangle=0 ~ \ . 
\end{equation}
Since $|\tilde{u}\rangle$ is orthogonal to all bound states including resonance anamneses, we can write: 
\begin{equation}\label{a2}
\hat{p}({\tilde e}-\hat{h})|\tilde{u}\rangle=0 ~ \ . 
\end{equation}
The general solution of Eq. (\ref{a1}) can be written as: 
\begin{equation}\label{a3}
|\tilde{u}\rangle=|\tilde{u}^{h}\rangle+\sum_{n}\alpha_{n}|\tilde{U}_{n}\rangle ~ \ ,
\end{equation}
where $|\tilde{u}^{h}\rangle$ is a solution of the homogeneous equation:
\begin{equation}\label{a4}
(\tilde{e}-\hat{h})|\tilde{u}^{h}\rangle=0 ~ \ , 
\end{equation}
$\{|\tilde{U}_{n}\rangle\}$ are solutions of the particular equations: 
\begin{equation}\label{a5}
(\tilde{e}-\hat{h})|\tilde{U}_{n}\rangle=|\tilde{u}_{n}\rangle ~ \ ,
\end{equation}
and $\{\alpha_{n}\}$ are constants. 
The orthogonality of $|\tilde{u}\rangle$ with respect to all bound states and anamneses of resonances:
\begin{equation}\label{a6}
\langle\tilde{u}_{m}|\tilde{u}\rangle=0 ~ \ , 
\end{equation}
can be expressed as: 
\begin{equation}\label{a7}
\vec{A}+B\vec{\alpha}=\vec{0} ~ \ ,
\end{equation}
where components of the vector $\vec{A}$ are given by: 
\begin{equation}\label{a8}
A_{m}=\langle\tilde{u}_{m}|\tilde{u}^{h}\rangle ~ \ , 
\end{equation}
and those of the matrix $B$ by: 
\begin{equation}\label{a9}
B_{mn}=\langle\tilde{u}_{m}|\tilde{U}_{n}\rangle ~ \ . 
\end{equation}
Inverting the linear system of equations (\ref{a7}), we obtain finally: 
\begin{equation}\label{a10}
|\tilde{u}\rangle=|\tilde{u}^{h}\rangle-\sum_{n,m}|\tilde{U}_{n}\rangle B^{-1}_{nm}A_{m} ~ \ . 
\end{equation}
The homogeneous solution has been  discussed in Sect. \ref{radicons}. Its  asymptotic behavior is given in Eq. (\ref{asym_scatt}). The states $\{|\tilde{U}_{n}\rangle\}$ are solutions of 
inhomogeneous equations with outgoing wave asymptotics: 
\begin{equation}\label{a11}
\tilde{U}_{n}(r)\sim C_{n}krh^{+}(kr) ~ \ , 
\end{equation}
where $\{C_{n}\}$ are constants. Inserting (\ref{asym_scatt}) and (\ref{a11}) into the radial representation of Eq.  (\ref{a10}), one derives the asymptotic behavior of the general solution: 
\begin{equation}\label{a12} 
\tilde{u}(r)=C^{-}i^{-(l+1)}\Big{[}(-1)^{l+1}e^{-i(kr-\eta\log(2kr))}+\tilde{S}e^{i(kr-\eta\log(2kr))}\Big{]} ~ \ , 
\end{equation}
where the scattering matrix is given by: 
\begin{equation}\label{a13}
\tilde{S}=\frac{1}{C^{-}}\Big{(}C^{+}-\sum_{n,m}C_{n}B^{-1}_{nm}A_{m}\Big{)}=e^{2i\tilde{\delta}} ~ \ , 
\end{equation}
and $\tilde{\delta}$ is the non-resonant phase shift. 

A possible test to see whether this procedure correctly removes the s.p. resonances from the 
s.p. scattering continuum is to compare the real-energy continuum phase shift for both Hamiltonians: $\hat{h}$ and $\hat{\tilde{h}}$. One demands that resonances are suppressed from the scattering continuum of $\hat{\tilde{h}}$. On the other hand, the background phase shift calculated for $\hat{h}$ and $\hat{\tilde{h}}$ should be essentially the same. 

\section{Discussion of the results}
\label{applications}
In this section, we shall discuss examples of  resonance anamneses constructed by the method described in Sects. \ref{radicons} and \ref{new_qbsec}. We shall also analyze properties of an associated non-resonant scattering continuum described in Sect. \ref{non_res_cont}. 

The starting point is the construction of an initial s.p. basis containing bound and scattering states. This basis is generated by the spherical  Woods-Saxon (WS) potential consisting of both central and spin-orbit parts, and a Coulomb potential: 
\begin{eqnarray}\label{potential}
V(r) &=& -V_0 f(r) - 4 V_{\rm so}~ (l \cdot s) \frac{1}{r} \frac{df(r)}{dr} + V_{\rm c}(r) ~ \ ,
\label{potential_1} 
\end{eqnarray}
where
\begin{eqnarray}
f(r) &=& \left[ 1 + \exp \left( \frac{r-R_0}{d} \right) \right]^{-1} ~ \ . \label{potential_2}
\end{eqnarray}
In all examples, the WS potential has the radius $R_0$=3.5 fm, the diffuseness 
$d=$0.5 fm, and the spin-orbit strength $V_{\rm so}$=3.5~MeV. The mass of a target nucleus is 16~amu. The Coulomb potential $V_{\rm c}$ is assumed to be generated by a uniformly charged sphere of radius $R_0$ and charge number $Z$=8. The depth of the central part is varied to simulate different situations.
The resonance anamneses are extracted from the scattering continuum of the WS potential (\ref{potential}). Bound states, anamneses of resonances and remaining continuum states are subsequently reorthogonalized.

\subsection{Radial wave functions of resonance anamneses and associated phase shifts} 
As an illustration of the method, we shall consider the $0d_{5/2}$ proton resonance in the WS potential (\ref{potential_1}) with $V_{0}$=40~MeV. The $0d_{5/2}$ resonance pole is situated at 
${\cal R}e(k^{res})=0.265\,\rm{fm}^{-1}$,  ${\cal I}m(k^{res})=-0.0014\,\rm{fm}^{-1}$, what corresponds to the energy $e^{res}$=1.55~MeV and the width $\Gamma^{res}$=32.692~keV. 
\begin{figure}[htbp]
\begin{center}
\includegraphics{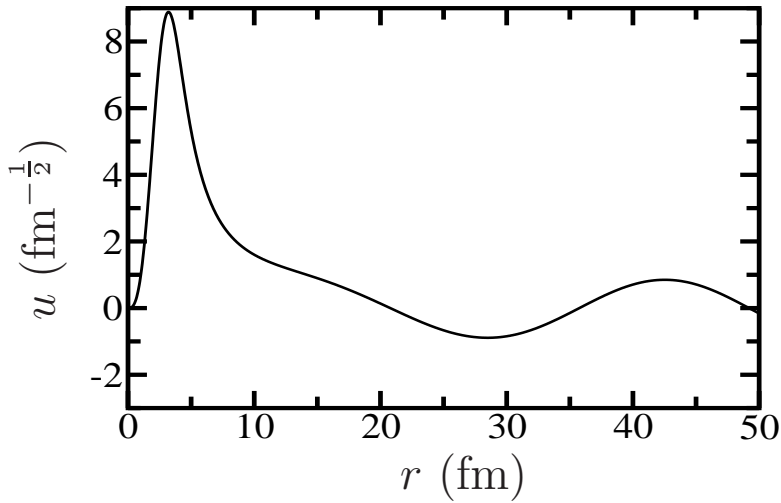}
\caption{The radial wave function of a $d_{5/2}$ proton scattering state at an energy corresponding to the real part of a $0d_{5/2}$ resonance state energy. For more details, see the description in the text. }
\label{scatt_0d5_proton}
\end{center}
\end{figure}
The radial wave function of a scattering state with real energy $e^{res}$ is presented in Fig. 
\ref{scatt_0d5_proton}. One can see the localization  of the wave function in the inner region (up to $\sim$10~fm). Solving the matching condition (\ref{qbsec_cont}) yields the matching radius: 
$r_{m}$($\equiv R$)=18~fm. 

\begin{figure}[htbp]
\begin{center}
\includegraphics{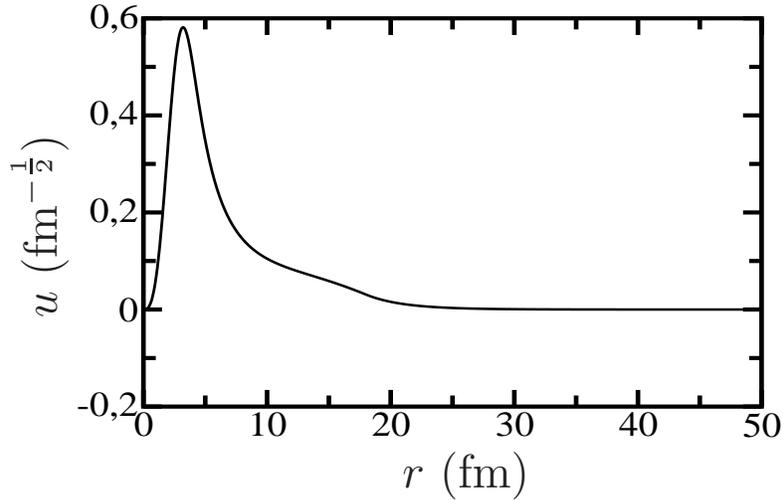}
\caption{The resonance anamnesis corresponding to the $d_{5/2}$ proton scattering state shown in Fig.  
\ref{scatt_0d5_proton}. }
\label{qbsec_0d5_proton}
\end{center}
\end{figure}
The resonance anamnesis corresponding to the $0d_{5/2}$ proton resonance (see Fig. 
\ref{qbsec_0d5_proton}) is extracted from the $d_{5/2}$ scattering wave function function at the energy $e^{res}$ corresponding to the real part of the $0d_{5/2}$ resonance energy (cf Fig. \ref{scatt_0d5_proton}). One may notice a hump at larger distances ($10~{\rm fm}\leq r \leq20~{\rm fm}$) which is a characteristic feature of many resonance anamneses.

\begin{figure}[htbp]
\begin{center}
\includegraphics{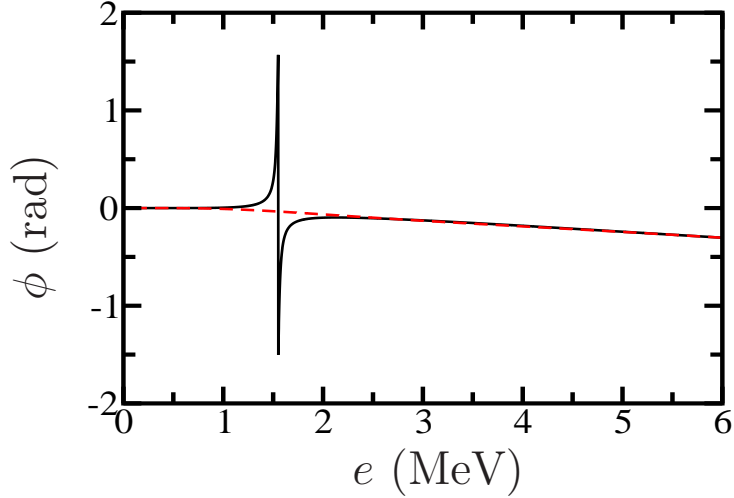}
\caption{Resonant (full line) and non-resonant (dashed line) $d_{5/2}$ phase shift as a function of the energy $e=\hbar^{2}k^{2}/2\mu$ in the center of mass. }
\label{phase_0d5_proton}
\end{center}
\end{figure}
$d_{5/2}$ phase shifts of the resonant continuum generated by the Hamiltonian $\hat{h}_{\rm WS}$ with the WS potential (\ref{potential_1}), and of the non-resonant continuum generated by the modified Hamiltonian $\tilde{\hat h}_{\rm WS}$ (cf Sect.  \ref{non_res_cont}), are plotted in Fig. \ref{phase_0d5_proton}. One can see that the resonant part of phase shift is fully removed by the $d_{5/2}$-resonance anamnesis and the phase shift calculated for the scattering continuum of  $\tilde{\hat h}_{\rm WS}$ exhibits only a smooth energy dependence. 

\begin{figure}[htbp]
\begin{center}
\includegraphics{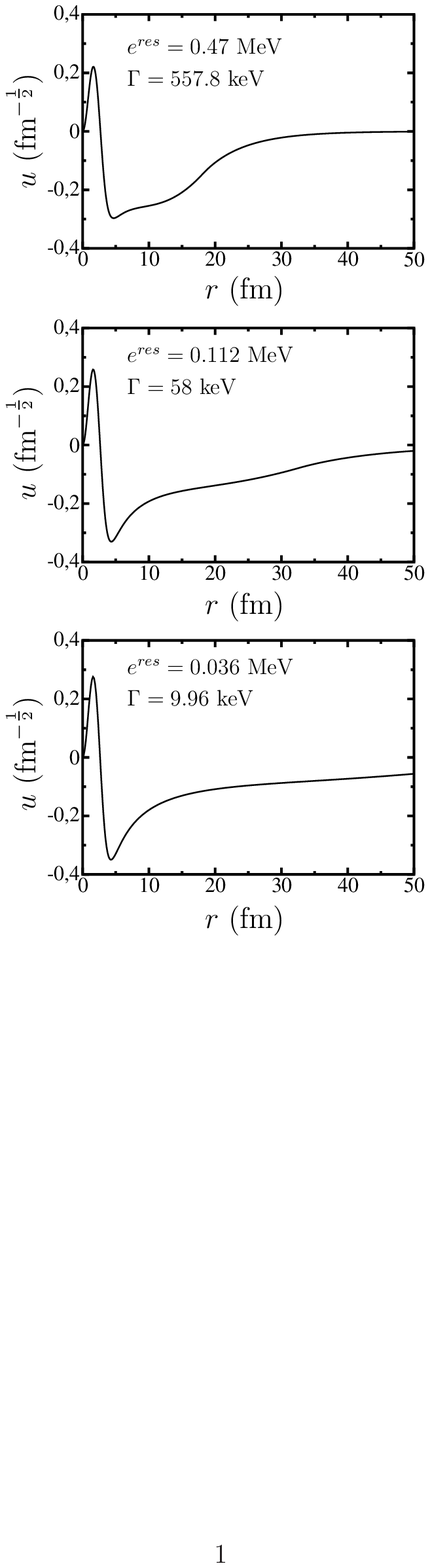}
\caption{Radial wave functions of resonance anamneses corresponding to $p_{1/2}$ neutron  scattering states at three different  energies of $1p_{1/2}$ neutron resonance. For more details, see the description in the text.}
\label{c3f4} 
\end{center}
\end{figure}
\begin{figure}[htbp]
\begin{center}
\includegraphics{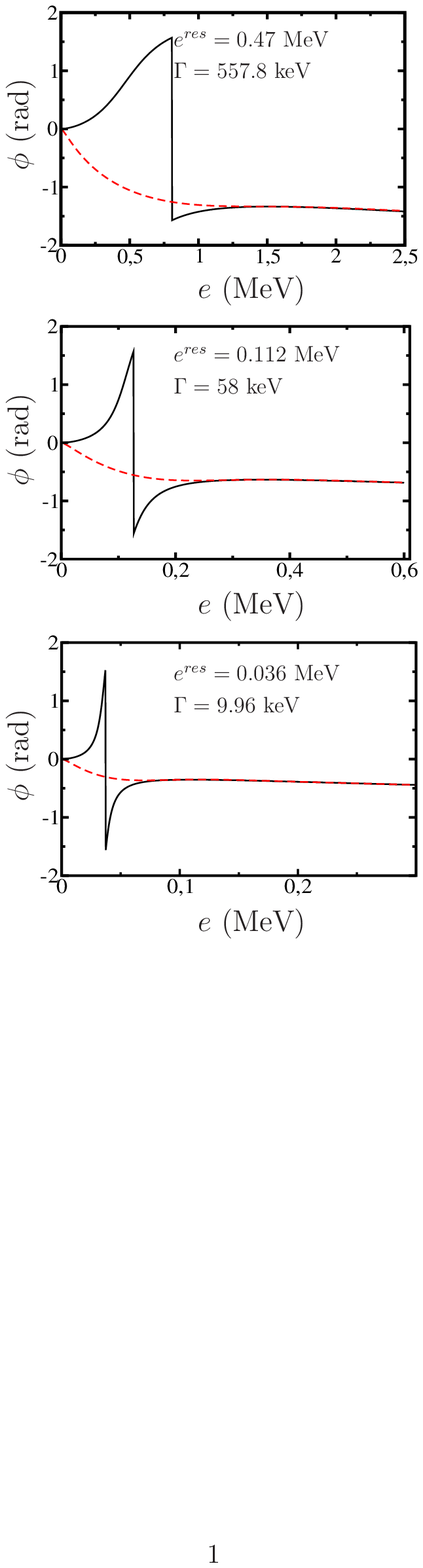}
\caption{Resonant (full line) and non-resonant (dashed line)  $p_{1/2}$ neutron phase shift for three resonances shown in Fig. \ref{c3f4}. }
\label{c3f5} 
\end{center}
\end{figure}
\begin{figure}[htbp]
\begin{center}
\includegraphics{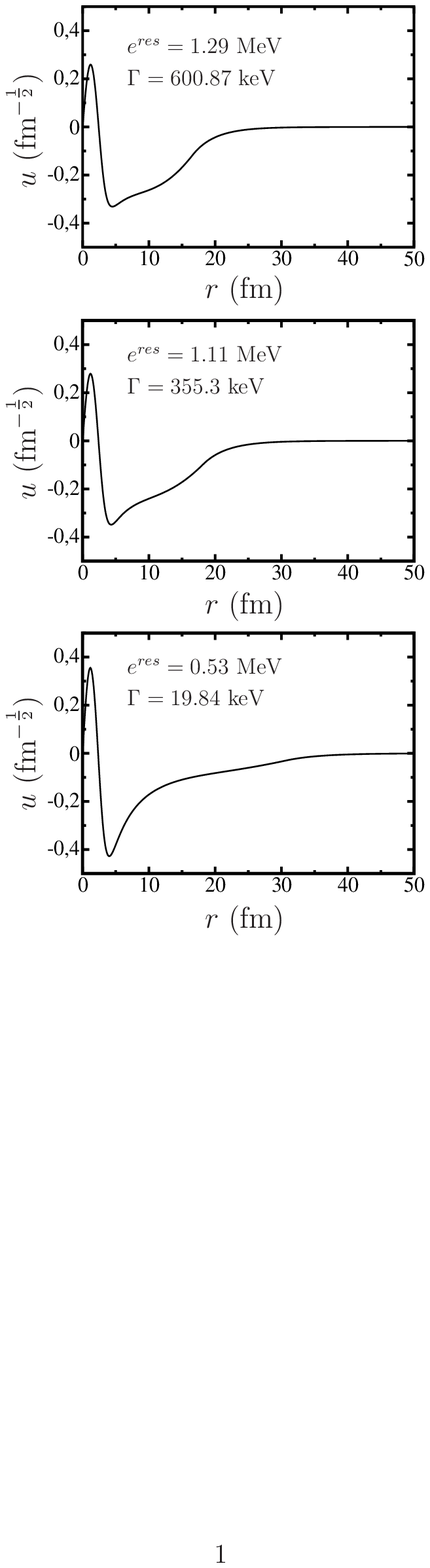}
\caption{Radial wave functions of resonance anamneses corresponding to $s_{1/2}$ proton scattering states at three different energies of $1s_{1/2}$ proton resonance. For more details, see the description in the text.}
\label{c3f8} 
\end{center}
\end{figure}
\begin{figure}[htbp]
\begin{center}
\includegraphics{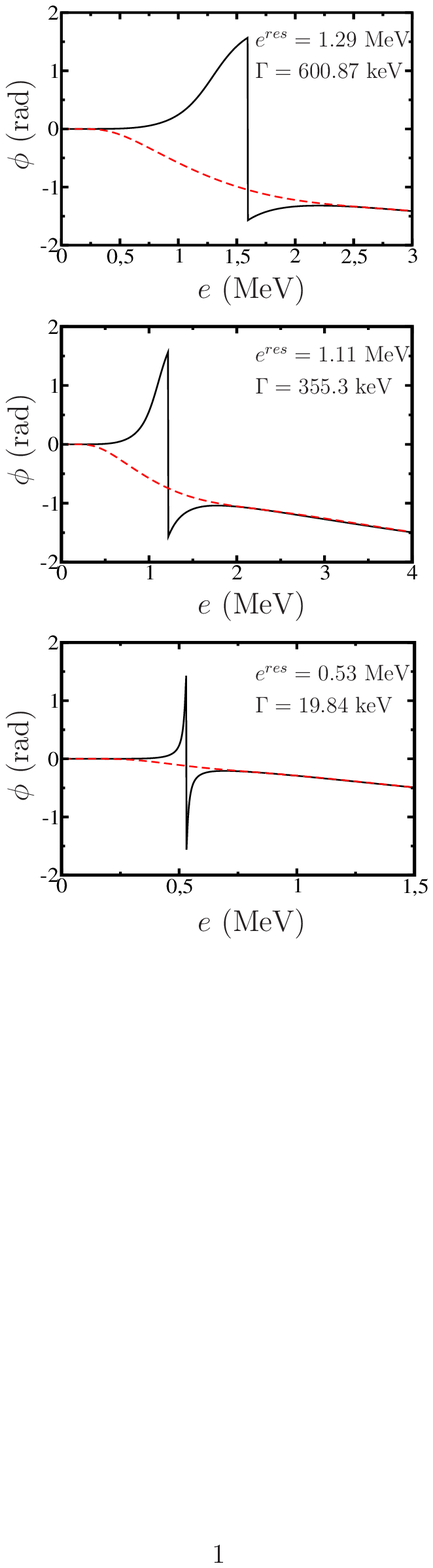}
\caption{Resonant (full line) and non-resonant (dashed line) $s_{1/2}$ proton phase shift for three resonances shown in  Fig. \ref{c3f8}. }
\label{c3f9} 
\end{center}
\end{figure}

Another examples of the radial wave functions of resonance anamneses and the respective phase shifts are shown in Figs. \ref{c3f4}-\ref{c3f9}. Fig. \ref{c3f4} shows the resonance-anamnesis wave functions corresponding to $1p_{1/2}$ neutron resonances at three different energies. The depth of the WS potential is varied to yield the resonance energy  $e^{res}=0.036$~MeV (the bottom part), 0.112~MeV (the middle part), and 0.47~MeV (the upper part). The widths of these resonances vary from $\sim10$~keV to $\sim560$~keV. The resonance anamneses are extracted from $p_{1/2}$ neutron scattering wave functions at energies $e^{res}$ corresponding to the real part of the $1p_{1/2}$ neutron resonance energy. In all these three cases, well pronounced hump at large distances can be seen.

Phase shifts in resonant  and non-resonant  scattering continua corresponding to those three $1p_{1/2}$ neutron resonances shown are plotted in Fig.   \ref{c3f5}. The resonant continuum is generated by the 
$\hat{h}_{\rm WS}$ Hamiltonian, whereas  the non-resonant continuum is given by the  modified Hamiltonian $\tilde{\hat h}_{\rm WS}$. One can see that the regularization procedure fully subtracts the resonant part of the phase shift, both for broad and narrow resonances, and the phase shift corresponding to the continuum of  $\tilde{\hat h}_{\rm WS}$ exhibits only a smooth energy dependence. 

Radial wave functions  of resonance anamneses and corresponding phase shifts for $1s_{1/2}$ proton resonance  at three different energies are shown in Figs. \ref{c3f8} and \ref{c3f9}. The depth of the WS potential is varied to yield the resonance energy  at $e^{res}=0.53$~MeV (the bottom part), 1.11~MeV (the middle part), and 1.29~MeV (the upper part). The resonance width in these cases vary from 
$\sim20$ keV to $\sim600$ keV. Also in this case, the extraction of the resonance anamnesis removes all resonant features from the scattering continuum of $\hat{h}_{\rm WS}$.

In all cases, the anamnesis of a resonance constructed by the method proposed in this work provides an excellent description of localized aspects of the resonant wave function in a broad range of resonance widths. The so defined resonance anamnesis becomes an image of the resonance wave function in the subspace of ${\cal L}^2$-functions. Its removal from the subspace of scattering wave functions leaves only the non-resonant continuum with smoothly varying phase shifts. This feature of resonance-anamnesis wave functions give them a special significance in a decomposition of the scattering continuum.

\subsection{RMS radius and matching radius of the resonance-anamnesis wave functions}
In the following, we shall investigate the energy dependence of the resonance-anamnesis radial wave functions by calculating the RMS radius and the matching radius. 

The dependence of the mean-square radius of a bound state on the distance $e$ to the continuum threshold has been studied by Riisager et al. \cite{Rii92} for spherical nuclei and Misu et al.  \cite{Mis97} for deformed nuclei.  (For the discussion of three-body halo asymptotics, see Ref. \cite{Fed93}.) The mean-square radius of a spherical neutron orbit varies as $(-e)^{-1}$ for $\ell=0$ and as $(-e)^{-1/2}$ for $\ell=1$, but remains finite for higher angular momenta due to the centrifugal barrier that confines the radial wave function. For proton bound states, the mean-square radius is finite for all $\ell$ at the continuum threshold because of the Coulomb barrier. 

The resonance-anamnesis wave functions constructed by the method proposed in this work have an asymptotic behavior of a bound state wave function with momentum $\sqrt{{\cal R}e\Big{(}(k^{res})^{2}\Big{)}}$. Consequently, it is likely that certain near-threshold features of resonance-anamnesis wave functions are similar to features of weakly bound s.p. states. In other words, one expects not only that the above method provides states which provide a faithful image of localized aspects of resonances, but also that it provides states which are a natural continuation of weakly bound s.p. states into the continuum. 

\begin{figure}[htbp]
\begin{center}
\includegraphics{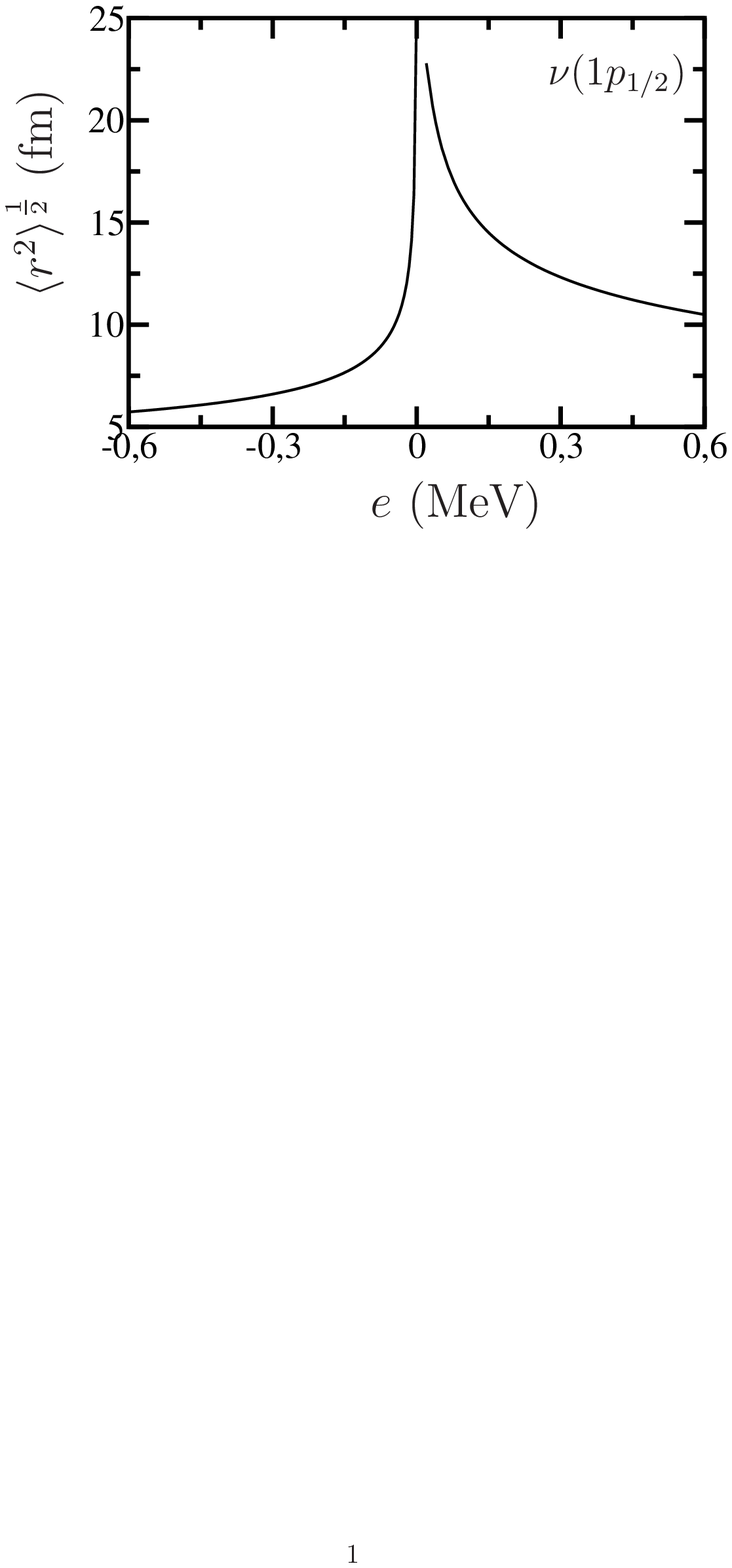}
\caption{RMS radius of $1p_{1/2}$ neutron orbit as a function of the energy. The branch for positive energies is obtained for $1p_{1/2}$ resonance anamnesis constructed by the method described in Sect. \ref{new_qbsec}. For more details, see the description in the text.}
\label{1p1n_msr}
\end{center}
\end{figure}
\begin{figure}[htbp]
\begin{center}
\includegraphics{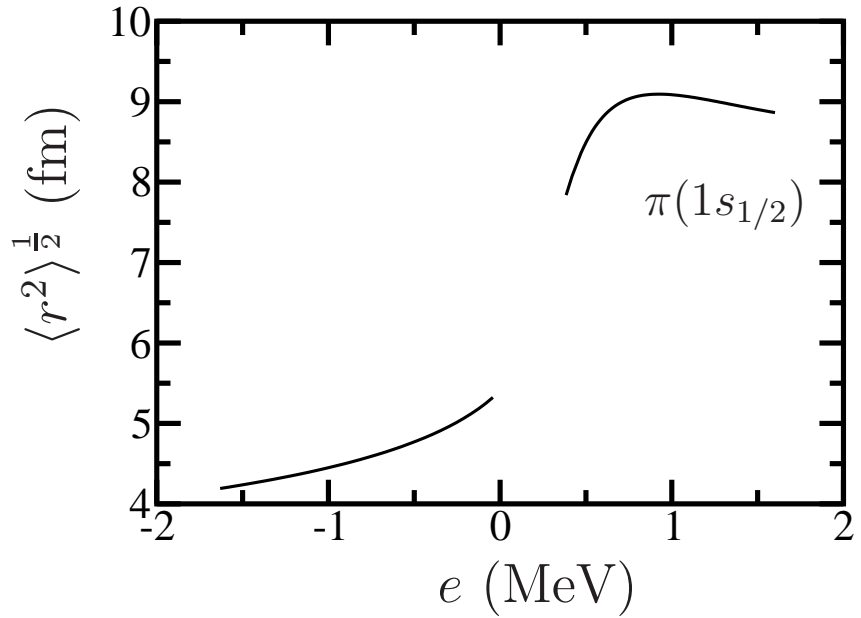}
\caption{RMS radius of the $1s_{1/2}$ neutron orbit as a function of the energy. The curve for positive energies is obtained for the $1s_{1/2}$ resonance anamnesis constructed by the method described in Sect. \ref{new_qbsec}. For more details, see the description in the text.}
\label{1s1p_msr}
\end{center}
\end{figure}
Figs. \ref{1p1n_msr} and \ref{1s1p_msr} present the RMS radius of a weakly bound s.p. state and a corresponding resonance anamnesis as a function of the energy for $1p_{1/2}$ neutron and $1s_{1/2}$ proton s.p. orbits, respectively. The curve for negative energies corresponds to the RMS radius for bound  s.p. states: neutron $1p_{1/2}$ (Fig.  \ref{1p1n_msr}) and proton $1s_{1/2}$ (Fig.  \ref{1s1p_msr}) states. For positive energies, the curve shows the RMS radius of the corresponding
resonance-anamnesis wave function extracted from the scattering wave  function at the energy which is equal to the real part of  the resonance energy, respectively. The corresponding bound and real-energy scattering states are found by varying  the depth of the WS potential (\ref{potential_1}). For  neutron $1p_{1/2}$ orbits  (cf Fig.  \ref{1p1n_msr}), the depth of the central part $V_0$ varies  from 69.012~MeV to 61.44~MeV, whereas for proton $1s_{1/2}$ orbits, $V_0$ changes from 50.233~MeV to 39.06~MeV. 
 
The gap seen in Figs. \ref{1p1n_msr} and \ref{1s1p_msr} close to $e=0$ comes essentially from numerical limitations in the construction of reliable wave functions for a weakly-bound state and its resonance anamnesis at positive energies. One may notice a slight asymmetry in the energy dependence of RMS radius on both sides of $e=0$. This is essentially an effect of the centrifugal and/or Coulomb barrier for states with $e>0$.

In all previous CSM or SMEC studies \cite{Ben99,Ben00,Oko03,Rot05}),  the QBSEC radial wave 
function was taken to be proportional to a regular solution of  Eq. (\ref{local_schr}) up to a certain radius $r_{cut}$, and for $r>r_{cut}$ this solution was cut by a cutting function. The value of  $r_{cut}$ in such an approach is a free parameter. For most of CSM/SMEC applications, the  value of $r_{cut}$ was close to the position of the top of the centrifugal and/or Coulomb barrier. 

\begin{figure}[htbp]
\begin{center}
\includegraphics{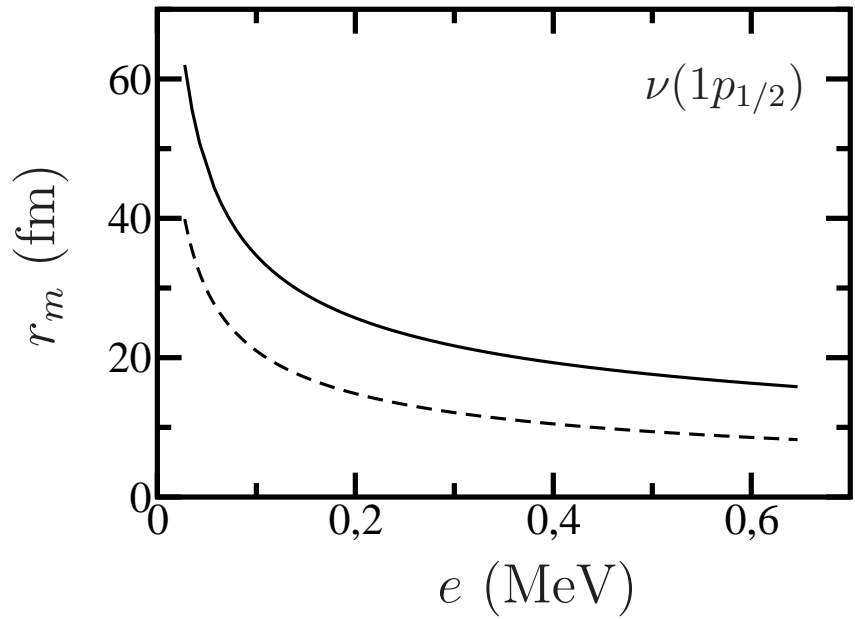}
\caption{Matching radius $r_{m}$ (full line) in function of the energy $e$ of the resonant 
$1p_{1/2}$ neutron state. The depth $V_{0}$ of the potential is varied from $66.1$ MeV to 
$61.44$ MeV. For comparison, the external turing point (dashed line) is presented too. }
\label{match_1p1n}
\end{center}
\end{figure}
As discussed in Sect. \ref{new_qbsec}, the resonance-anamnesis wave functions which replace QBSEC functions, are proportional to the regular solution of Eq. (\ref{local_schr}) up to the matching radius $r_{m}\equiv R$.  For $r>r_{m}$, the resonance anamnesis takes an asymptotic form of a bound state with a well defined wave number. The matching radius is not a free parameter of the theory but its value is uniquely determined by solving Eq. (\ref{qbsec_cont}) for $r$. At $r=R$, inner and outer solutions join smoothly to form a ${\cal C}(1)$ radial wave function of the resonance anamnesis. 

The matching radius $r_m$, whose position depends on features of the studied resonance, is an interesting supplementary information about the resonance-anamnesis wave function. $r_m$ depends uniquely on resonance characteristics, such as the energy (width), angular momentum and $\tau_z$. The energy (width) dependence of $r_m$ is generic, depending only  on the angular momentum and $\tau_z$. This dependence is illustrated in Fig. \ref{match_1p1n} on an example of  neutron $1p_{1/2}$ wave function of the resonance anamnesis. We can see that $r_m$ grows with decreasing $e$. The dashed line in  Fig. \ref{match_1p1n} shows also the variation of an external turning point $r_{etp}$ of the centrifugal barrier with the resonance energy. Except for a shift, both functions $r_m(e)$ and $r_{etp}(e)$ are rather similar, in particular for large $e$. In all studied cases, the matching radius lies outside of the external turning point of the potential barrier.

\section{Conclusions}
\label{conclusion}
Construction of the many-body basis in Hilbert space is a key problem in the SM description of open quantum systems, such as the weakly bound nuclei, radioactive decays, or low-energy nuclear reactions. This construction is based on the determination of a complete s.p. basis which consists of  discrete states (subspace $q$) and a scattering continuum (subspace $p$). A consistent unified formulation of nuclear structure and reactions is possible within the CSM/SMEC if the s.p. resonances are removed from the scattering continuum $p$ to be put in $q$ \cite{Bar77}. This regularization procedure for s.p. resonances is associated with an extraction of localized part of s.p. resonances from $p$ which leaves only non-resonant scattering states. 

In this paper, we have developed an unambiguous, parameter-free method of extracting the localized component of the s.p. resonance which allows to determine a new subspace $q^{'}$ of discrete s.p. states, consisting of bound states and resonance anamneses, and a new subspace $p^{'}$ of non-resonant scattering states. The resulting s.p. basis is complete and can be used for the construction of complete many-body basis in the Hilbert space. This new s.p. basis could also provide an interesting alternative for solving Hartree-Fock-Bogolyubov equations with finite-range density-dependent interactions for weakly-bound systems. 

The resonance anamneses provide a faithful image of localized features of resonances in the space of 
${\cal L}^2$-functions, removing all resonant aspects from calculated phase shifts. These anamneses of the resonance states provide also a smooth continuation of weakly-bound s.p. states into the low-energy continuum. Appealing features of those states give a special significance to the s.p. basis which includes them in its discrete part.  Future applications of this particular basis may open new and yet unexplored horizons for a direct calculation of multi-particle resonant states in nuclear systems. It may also be useful in extracting a resonance contribution from many-body observables in weakly-bound or unbound systems.

\vspace{1cm}
{\bf Acknowledgements:} \\
One of us (M.P.) wish to thank Witek Nazarewicz for useful comments.

\end{document}